\documentclass[prb,twocolumn, floatfix]{revtex4}
\usepackage{graphicx}
\usepackage{epstopdf}
\usepackage{bm, amsmath, amssymb}

\newcommand{\nbar}[0]{\bar{n}}

\newcommand{\qext}[0]{Q_\textrm{\small{ext}}}

\begin{document}

\title{1/f noise of Josephson-junction-embedded microwave resonators at single photon energies and millikelvin temperatures}
\author{K. W. Murch} \email[Corresponding author: ]{katerm@berkeley.edu}
\author{S. J. Weber}
\author{E. M. Levenson-Falk}
\author{R. Vijay}
\author{I. Siddiqi}
\affiliation{Quantum Nanoelectronics Laboratory, Department of Physics, University of California, Berkeley CA 94720}

\date{\today}

\begin{abstract}
We present measurements of 1/f frequency noise in both linear and Josephson-junction-embedded superconducting aluminum resonators in the low power, low temperature regime\textemdash typical operating conditions for superconducting qubits.  The addition of the Josephson junction does not result in additional frequency noise, thereby placing an upper limit for fractional critical current fluctuations of $1 \times 10^{-8}$ ($1/\sqrt{\textrm{Hz}}$) at 1 Hz for sub-micron, shadow evaporated junctions. These values imply a minimum dephasing time for a superconducting qubit due to critical current noise of 40 -- 1400 $ \mu$s depending on qubit architecture.  Occasionally, at temperatures above 50 mK, we observe the activation of individual fluctuators which increase the level of noise significantly and exhibit Lorentzian spectra.    \end{abstract}

\maketitle

Superconducting quantum electronics rely on low-loss inductive, capacitive, and nonlinear elements to achieve long-lived coherence in quantum circuits and minimal added noise in analog detectors and amplifiers.  Progress has been made in recent years to mitigate high frequency loss in superconducting circuits, improving the energy relaxation time of superconducting qubits and the quality factor of superconducting resonators\cite{wang09,bare10loss}.  The frequency stability of such resonant circuits is limited by low frequency temporal fluctuations of material properties, which typically obey a universal 1/f$^\alpha$ power spectrum with $\alpha \sim 1$. These fluctuations currently limit the sensitivity of astrophysical sensors\cite{day03} and the coherence time of superconducting qubits\cite{mart03Decoherence,vanh04,paik113D} with slow energy relaxation. Such ``1/f noise'' may arise from defect induced dielectric and magnetic fluctuations as well as fluctuations of the critical current in structures incorporating Josephson junctions. 
Frequency noise in linear, GHz-frequency superconducting resonators has been studied at high excitation power to improve the noise limits on microwave kinetic inductance detectors\cite{day03,gao08,bare10,gao11}. In tunnel junctions, critical current noise has been inferred from fluctuations of the normal state resistance in Al, In, Pb and Nb two-junction SQUIDs\cite{vanh04,erom06lowfreq,pott09temp,vanh11com}.  



In this letter, we present measurements of 1/f frequency fluctuations in 4-8 GHz coplanar waveguide (CPW) and lumped element (LE) aluminum superconducting resonators at $T=25$ mK and excitation powers down to the single photon regime.  Furthermore, we incorporate Josephson tunnel junctions with areas up to 0.44 $\mu$m$^2$ and $I_0\sim 1\ \mu$A into these resonators and measure the resulting phase noise to extract the frequency dependence and intensity of junction inductance fluctuations in the zero-voltage state under weak excitation. Typically, the junction embedded resonators exhibit no observable noise above the intrinsic fluctuations of the linear resonator. These measurements thus set an upper bound for fractional critical current fluctuations, $S_I^{1/2}< 1\times10^{-8}$ ($1/\sqrt \textrm{Hz}$) at 1 Hz in sub-micron, shadow evaporated aluminum junctions. This value is one order of magnitude lower that previous measurements \cite{erom06lowfreq,pott09temp,vanh11com} and suggests that these junctions are suitable for quantum circuits with coherence times up to $\sim1$ ms. 

Our CPW resonators were fabricated in a liftoff process from aluminum films deposited on high-resistivity silicon and consisted of a capacitivley isolated section of 50 $\Omega$ transmission line with a center trace width of 20 $\mu$m. 
The LE resonators employ $\sim$ pF parallel plate capacitors with a single crystal silicon dielectric\cite{webe11} shunted with a meander inductor. 
Josephson-junction-embedded resonators were formed from a double-angle evaporation process and consisted of either a single junction or two junctions in a SQUID geometry shunted with $C = 1.1$ pF.  
For each junction embedded resonator, we co-fabricated a linear resonator with the same metal deposition/oxidation sequence used to produce the tunnel junctions. This allows us to isolate the contribution of critical current fluctuations from other sources of noise.  The parameters of the Josephson-junction-embedded samples and a typical co-fabricated linear resonator are given in Table I.

\begin{figure}
\includegraphics[angle = 0, width = 0.5\textwidth]{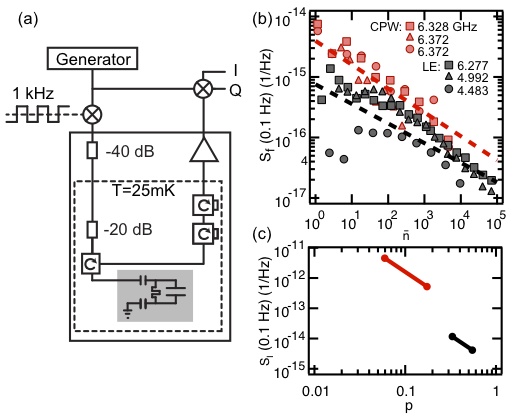}
\caption{\label{fig3}  (a) Measurement setup. (b) The spectral density of frequency fluctuations at 0.1 Hz as a function of internal resonator photon number, $\nbar$, for CPW  and LE resonators.  (c) The critical current noise sensitivity for  CPW (red) and LE (black) embedding geometries as a function of the Josephson inductance participation ratio. }
\end{figure}

Samples were thermally anchored to the mixing chamber stage of a dilution refrigerator and cooled to $T=25$ mK.  The samples were shielded by successive layers of superconducting and magnetic shields and probed via heavily attenuated coaxial lines.  A schematic of the measurement setup is shown in Figure 1(a).   The phase of the transmitted or reflected microwave tone was measured using homodyne detection and digitized at 100 MS/s.  The relative phase between the local oscillator and probe tone was chopped at 1 kHz and synchronously detected to reduce the effect of 1/f noise in the offsets of room temperature components.  


To examine the frequency noise of the samples, multiple records of the reflected or transmitted microwave signal phase were acquired for 30 s.   We calibrated the phase noise floor of our measurement chain by examining the RMS spectral density of phase fluctuations $S_{\theta}$  of the reflected microwave signal at a drive frequency away from the resonance where all the incident power was reflected.   The phase noise floor was dominated by white noise of the amplification chain for frequencies above 1 Hz.   At lower frequencies, a 1/f technical noise floor of $S_\theta^\textrm{tech} = -85 $ dBc/Hz at 0.1 Hz was observed.  In all cases, we subtract the white and technical noise components from the phase noise.  When probed on resonance, the variation of the phase with the resonator frequency can be linearized: $d\theta/df  \approx \pi Q/f_0$. As such, when biased in this regime, any fluctuations of the resonator frequency appear as phase fluctuations of the homodyne output signal.   The spectral density of fractional frequency fluctuations was determined from the phase noise and the phase response of the resonance, $S_f \equiv  S_{\theta}/(f_0d\theta/df)^2$. 
 
In our measurement technique, the resonator quality factor sets lower and upper limits on the magnitude of frequency noise that can be observed.  The maximum noise intensity that can be measured by monitoring the phase of a high-$Q$ resonance is $S_f^\textrm{max} \approx 1/(\pi Q)^2$; a high quality factor diminishes the dynamic range of detectable frequency excursions.  The minimum resolvable noise intensity is set by the technical phase noise floor, $S_f^\textrm{min} \simeq S_\theta^\textrm{tech}/(\pi Q)^2$.  We choose the coupled resonator $Q$ to be between $10^3$ and $10^4$ to resolve the intrinsic resonator fluctuations above the noise floor, with sufficient dynamical range to encompass the highest levels of noise observed.
 

\begin{table}[htdp]
\caption{Sample parameters.  Sample A was a linear LE resonator.  Sample D incorporated a single junction rather than 2 junctions in a SQUID geometry. }
\begin{center}
\begin{tabular}{l c c c  ccccc }
      \hline
\hline
Sample&&$f_0$(GHz)&&$p\equiv L_J/L_\textrm{tot}$ &&$I_0\ (\mu$A)& &$J_c$ (A/cm$^2$)\\
\hline
A & 	&6.34&$\ $	&0		&$\ $&-		&$\ $&-\\
B &  		&6.45			&&0.50			&&1.2	&&270\\
C &  		&5.41			&&0.55			&&1.2	&&270\\
D &		&6.38			&&0.52			&&1.1	&&390\\
E &  		&7.20			&&0.40			&&1.8	&&1200\\
F & 			&7.68			&&0.33			&&2.5	&&1200\\
\hline
\hline
\end{tabular}
\end{center}
\label{default}
\end{table}%
 
In Figure 1(b), we show $S_f(0.1\ \textrm{Hz})$ for several linear CPW and LE resonators as a function of excitation power down to average photon number $\nbar = 1$. Here, $\nbar$ is related to the drive power as $\nbar=2 P_\textrm{\small{in}} Q^2/(\pi m \qext h f_0^2)$ on resonance, where $m=1$ for the LE resonators and $m=2$ for the CPW resonators. The total quality factor $Q=( 1/Q_i + 1/\qext)^{-1}$ was limited by deliberate coupling to the environment.    The measured frequency noise decreased with increasing resonator excitation as $\nbar^{-\beta}$, where $\beta = 0.32-39$, similar to the scaling observed at higher power\cite{bare10}. 

These linear resonator samples can serve as a test vehicle for characterizing the 1/f critical current fluctuations in Josephson tunnel junctions. For a junction-embedded linear resonator,  assuming  that the observed frequency noise originates solely from junction inductance fluctuations, which can be expressed as critical current noise, we can place an upper bound on this noise. Fractional critical current noise is related to the observed frequency noise as $S_I \equiv  4 S_{f}/p^2$, where $p = L_J/L_\textrm{tot}$ is the participation of the junction inductance $L_J$ in the total inductance $L_\textrm{tot} = L+L_J$.  In Figure 1(c) we estimate the potential sensitivity of the frequency noise measurement to the critical current fluctuations of an embedded Josephson junction. We use the frequency noise values observed in our linear resonators to determine the minimum level of critical current noise that can be resolved in $\sim 1\ \mu$A junctions embedded in 4-8 GHz CPW and LE resonators for practically accessible values of $p$. 
Our LE resonators formed from single crystal silicon capacitors allow high participation of the junction inductance in the circuit and exhibit a lower level of intrinsic frequency noise, reducing the measurement limit for critical current noise.  For the junctions considered here in the LE geometry, the measurement sensitivity is below $10^{-26} $A$^2$/Hz, roughly 2 orders of magnitude lower than what has been resolved by measurements in the voltage state\cite{vanh11com}.

\begin{figure}
\includegraphics[angle = 0, width = 0.5\textwidth]{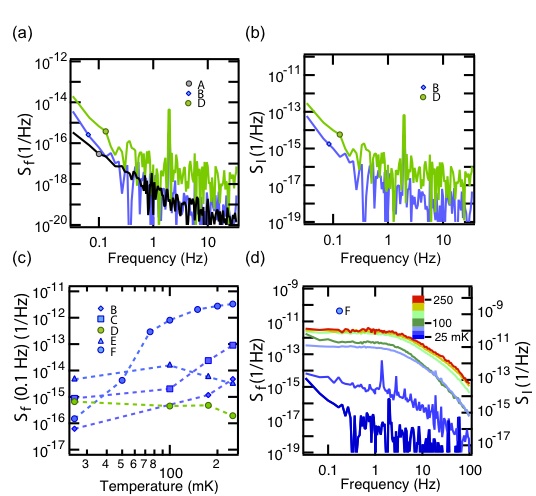}
\caption{\label{fig3} 1/f frequency noise of Josephson junction embedded resonators.  (a) The measured frequency  noise of junction embedded samples and the co-fabricated  linear resonator. (b) The measured critical current noise assuming all the frequency noise is attributed to critical current fluctuations of the junction.  (c) Temperature dependence of the junction-embedded resonators.  (d) The spectral density of frequency noise (left axis) and equivalent critical current noise noise (right axis) for sample F at temperatures up to 250 mK.}
\end{figure}

In Figure 2(a) we show the measured frequency noise of a few Josephson junction embedded resonators and a typical co-fabricated linear resonator at excitation powers that correspond to $\nbar\approx 1$ and zero flux threading the SQUID loop.  For samples B and D, the frequency noise was not significantly higher than what we observed for sample A, the co-fabricated linear resonator, allowing us to place only an upper limit on the level of critical current noise of the Josephson junctions. Furthermore, the noise was not significantly different between the SQUID and single junction samples, indicating that flux noise does not contribute to the measured noise.   We convert the observed frequency noise to an inferred critical current noise in Figure 2(b). 

In Figure 2(c) we plot the temperature dependence of the frequency noise at 0.1 Hz measured in the junction samples at $\nbar\approx1$. With the exception of sample F, the samples do not indicate a strong temperature dependence.   In Figure 2(d), we plot the spectral density of frequency noise of sample F as the temperature of the sample was raised incrementally to 250 mK. As the temperature of the sample was increased, the level of noise increased dramatically and the frequency response exhibited a single pole roll off. This level of noise was readily resolved above the frequency noise floor of the linear resonator. The response is characteristic of the activation of a single fluctuator above $T=50$ mK.   It is conceivable that an ensemble of such fluctuators could give rise to a 1/f noise spectrum with an intensity that would agree with voltage state measurements\cite{erom06lowfreq,pott09temp,vanh11com}.   

In conclusion, we have used a dispersive measurement technique to characterize the frequency noise of 4-8 GHz CPW and LE resonators, both with and without embedded Josephson junctions, in the single photon, low temperature regime.  Surprisingly, the noise intrinsic to the resonator structure dominates over the 1/f critical current noise at low temperatures. Using the observed level of fluctuations to set an upper bound on critical current fluctuations, and based on the analysis in Ref.\ (5), we estimate a lower limit on the dephasing time for 6 GHz phase, flux, and charge (transmon) qubits to be $T_2^*>$ 40, 80 and 1400 $\mu$s respectively. This level of observed noise is insufficient to account for the dephasing rates currently observed in superconducting qubits  and suggests that further performance improvements are in principle possible.

This research was funded by the Office of the Director of National Intelligence (ODNI), Intelligence Advanced Research Projects Activity (IARPA), through the Army Research Office. All statements of fact, opinion or conclusions contained herein are those of the authors and should not be construed as representing the official views or policies of IARPA, the ODNI, or the US Government.  The ARO QCT (R.V.) and DARPA YFA (E.M.L.-F.) programs  are acknowledged for proving financial support.



\end{document}